\documentclass[12pt]{iopart}
\usepackage{bm}
\usepackage{graphicx}
\usepackage{iopams}

\begin{document}

\title{`C$_{60}$ spin-charging' with an eye on a quantum computer}

\author{J P Connerade\S\
and V K Dolmatov\P}

\address{\S\ The Blackett Laboratory, Imperial College, Prince Consort
Road, London SW7 2BZ, UK E-mail: Jean-Patrick@Connerade.com}

\address{\P\ Department of Physics and Earth Science,
University of North Alabama, Florence, AL 35632, USA. E-mail: vkdolmatov@una.edu}

\begin{abstract}
A question whether there exists an interaction between the spins of the endohedral atom $A$@C$_{60}$ and the properties of the confining shell which might affect the alignment of, or manipulation by,
the spins for building a register for a quantum computer  is discussed. It is argued that an effect, termed the `C$_{60}$ spin-charging' effect, can occur in endohedral atoms and would affect the operation of a quantum register.
The effect is exemplified by choosing the $\rm 3d$ (Cr and Mn) and $\rm 4d$ (Mo and Tc) transition metal atoms as well as a rare-earth Eu atom as the case study. A class of high-spin atoms which are less suitable for building a quantum register is, thus, identified.
\end{abstract}

\submitto{\jpb}
\maketitle

The use of a non-zero spin atom confined by C$_{60}$ (referred to as the endohedral $A$@C$_{60}$ atom) as the building block of the register for a quantum computer
was proposed by Harneit \cite{Harneit}. Obviously, the higher the spin of the atom, the better. Confined atoms then must be atoms with one or more multielectron semifilled subshells in their configuration
whose electron spins are aligned. The study by Harneit \cite{Harneit} focused on the use of a semifilled shell N($\rm 2p^{3})@$C$_{60}$ atom.
The general idea for building the register for a quantum computer depends
on the freedom to align the spin of the
encapsulated atom, on the ability of the C$_{60}$ confining cage to screen the spins from the influence of unwanted
decohering fields and on the ability to write (read) to (from) an assembly of confined atoms held together as
an array.

It is, therefore, interesting to explore whether, in fact, the freedom to align the spins of encapsulated atoms exists independently
of the properties of a confining shell and  whether external fields are able to perturb this alignment. The latter question has already been
addressed theoretically by Connerade and Solovyov \cite{C&S} and Amusia and Baltenkov \cite{AmBalt} who studied the properties of a spherical
C$_{60}$ cage and showed under what conditions the C$_{60}$ screening of an external field remains effective. The former of the two questions
is addressed in the present paper by accounting for an effect termed the `C$_{60}$ spin-charging effect'.

The  C$_{60}$ spin-charging effect was recently uncovered as a by-product by Dolmatov \etal \cite{e+C60} in the study of ${\rm e^{-}} + A@{\rm{C}}_{60}$ electron elastic scattering.
The quintessence of the effect is that both the  electron spin-up $P_{\rm n\ell\uparrow}(r)$ and spin-down $P_{\rm n\ell\downarrow}(r)$ functions of a
high spin encapsulated atom $A$, such as an atom with one or more multielectron semifilled subshells in its configuration, may be drawn noticeably, but very differently,
into the region of the C$_{60}$ wall.
This results in loading the C$_{60}$ cage with electron density of a preferred spin orientation. Naturally, the effect is accompanied by the loss of some electron spin-density localized on
the confined atom $A$ itself. Clearly, the phenomenon is potentially important for the proposed realization of an  $A$@C$_{60}$ register for a quantum computer.
It is the ultimate aim of the present paper to delineate the spin-charging effect more precisely for this purpose. To meet this goal, the $\rm 3d$-, $\rm 4d$- and $\rm 4f$-semifilled shell
Cr(...$\rm 3d^{5}4s^{1}$, $^{7}\rm S$), Mn(...$\rm 3d^{5}4s^{2}$, $^{6}\rm S)$, Mo(...$\rm 4d^{5}5s^{1}$, $^{7}\rm S$), Tc(...$\rm 4d^{5}5s^{2}$, $^{6}\rm S$) and Eu(...$\rm 4f^{7}6s^{2}$, $^{8}\rm S$) atoms
encapsulated inside C$_{60}$ are chosen for the completeness of the case study.

In the following, we briefly outline the methodology to calculate the C$_{60}$ spin-charging effect in an endohedral semifilled shell atom, $A$@C$_{60}$.

A convenient way to account for
the structure of a semifilled shell atom is provided by the spin-polarized Hartree-Fock approximation (SPHF) developed by Slater \cite{Slater}.
 SPHF accounts for the fact that spins of all
electrons in a semifilled subshell of the atom (e.g., in the $\rm 3d^{5}$ subshell of Mn) are co-directed, in accordance with Hund's rule, say, all pointing upward. This results in splitting of
a closed ${n\ell}^{2(2\ell+1)}$ subshell in the atom into two semifilled subshells of opposite spin orientations, ${n\ell}^{2\ell+1}$$\uparrow$ and ${n\ell}^{2\ell+1}$$\downarrow$. This is in view of
 the presence of
exchange interaction between $n\ell$$\uparrow$ electrons with  only spin-up electrons in the original semifilled
subshell of the atom (like the $\rm 3d^{5}$$\uparrow$ subshell in the Mn atom) but  absence of such for $n\ell$$\downarrow$ electrons. Thus, the SPHF configurations of Cr, Mn, Mo, Tc and Eu are as
follows:
Cr(...${3\rm p}^{3}$$\uparrow$${3\rm p}^{3}$$\downarrow$${3\rm d}^{5}$$\uparrow$${4\rm s}^{1}$$\uparrow$, $^{7}$S),
 Mn(...${3\rm p}^{3}$$\uparrow$${3\rm p}^{3}$$\downarrow$${3\rm d}^{5}$$\uparrow$${4\rm s}^{1}$$\uparrow$$4s^{1}$$\downarrow$, $^{6}$S),
 Mo(...${4\rm p}^{3}$$\uparrow$${4\rm p}^{3}$$\downarrow$${4\rm d}^{5}$$\uparrow$${5\rm s}^{1}$$\uparrow$, $^{7}$S),
Tc(...${4\rm p}^{3}$$\uparrow$${4\rm p}^{3}$$\downarrow$${4\rm d}^{5}$$\uparrow$${5\rm s}^{1}$$\uparrow$${5\rm s}^{1}$$\downarrow$, $^{6}$S)
  and
Eu(...${4\rm d}^{5}$$\uparrow$${4\rm d}^{5}$$\downarrow$${4\rm f}^{7}$$\uparrow$...${6\rm s}^{1}$$\uparrow$$6s^{1}$$\downarrow$, $^{8}$S).
SPHF equations for the ground  state of a semifilled shell atom differ from ordinary HF equations for closed shell atoms by accounting for exchange interaction only between electrons with the same spin orientation
($\uparrow$, $\uparrow$ or $\downarrow$, $\downarrow$) \cite{Slater,ATOM}. To model a $A$@C$_{60}$ atom, we account for the presence of the C$_{60}$ cage by adding a rectangular (in the radial coordinate $r$) potential
well $U_{\rm C_{60}}(r)$ of a finite width $\Delta$, depth $U_{0}$ and inner radius $r_{0}$ to the HF (SPHF) equations \cite{@Ca}, as in many of other studies,
see, e.g., \cite{e+C60,JPCS,Pushka,McKoy,Phaneuf} and references therein:

\begin{eqnarray}
U_{\rm C_{60}}(r)=\left\{\matrix {
-U_{0}, & \mbox{if $r_{0} \le r \le r_{0}+\Delta$} \nonumber \\
0 & \mbox{otherwise.} } \right.
\label{SWP}
\end{eqnarray}

\begin{figure}[h]
\center{\includegraphics[width=10cm]{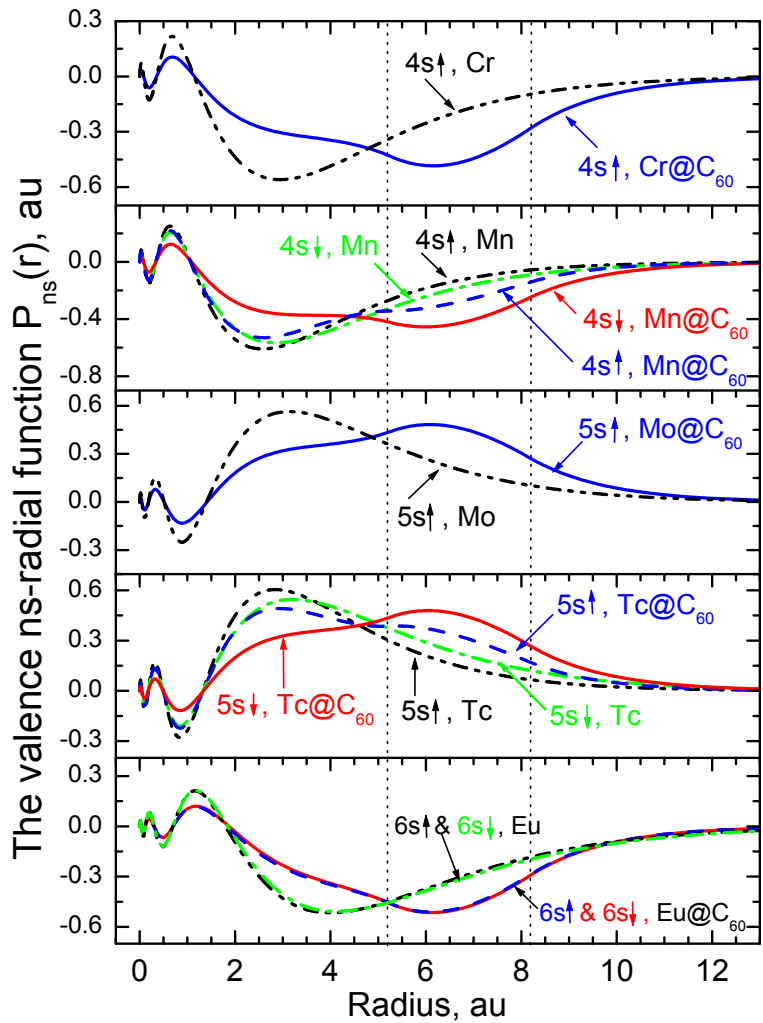}}
\caption{(Color online) Calculated $P_{n\rm s\uparrow}(r)$ and $P_{n\rm s\downarrow}(r)$ radial functions (in atomic units, au) of the valence subshells of the Cr@C$_{60}$,
Mn@C$_{60}$, Mo@C$_{60}$, Tc@C$_{60}$ and Eu@C$_{60}$ atoms versus those of the free atoms, as marked. Vertical dotted lines locate the position  of the C$_{60}$ wall,
$5.262 \le r \le 8.17$ au.}
\label{fig1}
\end{figure}

In the literature, some inconsistency is present in choosing the magnitudes of $\Delta$, $U_{0}$ and $r_{0}$ of the C$_{60}$ phenomenological potential (\ref{SWP}), cp., e.g., References
\cite{@Ca,e+C60,JPCS,Pushka,McKoy,Phaneuf} with each other. In the present paper, following \cite{McKoy}, we choose
$\Delta = 2.9102$ au (which is twice of the covalent radius of carbon), $r_{0} = 5.262$ au $ = R_{\rm c} - 1/2\Delta$ ($R_{\rm c}=6.7173$ au being the radius of the C$_{60}$ skeleton)
and $U_{0} = 7.0725$ eV (which was found  by matching the electron affinity $EA=-2.65$ eV of C$_{60}$ with the assumption that the orbital momentum of the $2.65$-eV-state is $\ell =1$ \cite{McKoy}). This choice is most
consistent with observations.
Calculated $P_{n\rm s\uparrow}(r)$ and $P_{n\rm s\downarrow}(r)$ functions of valence spin-up and spin-down electrons of the Cr, Mn, Mo, Tc and Eu atoms, both free and encapsulated inside C$_{60}$,
are depicted in figure \ref{fig1}.

Note how the encapsulation of the chosen atoms inside the C$_{60}$ cage draws their outer $P_{n\rm s\uparrow}(r)$ and, respectively, $P_{n\rm s\downarrow}(r)$ orbital functions
 into the region of the C$_{60}$ wall. This implies a significant transfer of electron density  from the encapsulated atom to the cage, but, more importantly in the context of the present paper,
a transfer of electron \textit{spin-density} from the atom to the cage. The transfer makes the cage become `spin-charged'. The C$_{60}$
cage becomes spin-charged even for the \textit{spin-neutral} $\rm 4s^{2}$ and $\rm 5s^{2}$  subshells of endohedral Mn and Tc, respectively. This is because of the stronger drain of the
valence $n\rm s$$\downarrow$ than
 $n\rm s$$\uparrow$
 electron density from the atom to the  cage. Interestingly enough, the spin-dependent drain of the valence electron density does not emerge in Eu@C$_{60}$ where both the
$P_{\rm 6s\uparrow}(r)$ and $P_{\rm 6s\downarrow}(r)$ orbital functions are drawn into the C$_{60}$ cage equally strongly, in contrast to the outer $P_{n\rm s\uparrow}(r)$ and $P_{n\rm s\downarrow}(r)$
orbital functions of the endohedral Mn and Tc atoms. This is because the semifilled $\rm 4f^{7}$$\uparrow$ subshell of Eu lies much deeper relative to its $\rm 6s^{1}$$\uparrow$ and $\rm 6s^{1}$$\downarrow$ subshells than the
semifilled $n\rm d^{5}$$\uparrow$ subshell of Mn ($n=3$) or
Tc ($n=4$) relative their valence $(n+1)\rm s^{1}$$\uparrow$ and $(n+1)\rm s^{1}$$\downarrow$ subshells.  For this reason, the exchange interaction between the $\rm 4f$$\uparrow$ and $\rm 6s$$\uparrow$ electrons in Eu is negligibly small. Hence,
there is practically no difference between the $P_{\rm 6s\uparrow}(r)$ and $P_{\rm 6s\downarrow}(r)$ functions both in free and encapsulated Eu. As a result, the Eu atom retains its electron \textit{spin-density}
intact upon confinement inside C$_{60}$, which could prove important for an eventual application.

In conclusion, the authors believe that the C$_{60}$ spin-charging effect we have described will affect the manipulation of spins in the corresponding $A$@C$_{60}$ systems and that it must inhibit, or at least render more complex, the operation of a quantum register. The present paper thus brings to light a class of high-spin atoms which are less suitable for building a quantum register, namely those which are subject to a strong electron
spin-density drain from the atom to the C$_{60}$ cage.
\ack
VKD acknowledges the supported of NSF Grant no. PHY-1305085. The undergraduate students C. J. Bayens, M. B. Cooper and M. E. Hunter of the University of North Alabama are thanked for their assistance with the calculations.
\section*{References}

\end{document}